\documentstyle[12pt]{article}
\begin{document}

\begin{titlepage}
\pagestyle{empty}
\title{On search for the M-Theory Lagrangian}
\author{Ioannis P.\ ZOIS\thanks{izois\,@\,maths.ox.ac.uk} \\
\\
Mathematical Institute,\\24--29 St.\ Giles', Oxford OX1 3LB, UK.}
\date{}
\maketitle
\begin{abstract}
We present a starting point for the search for a Lagrangian density for M-Theory using characteristic
classes for flat foliations of bundles.
\end{abstract}
\end{titlepage}

\section{}

 The starting point of our investigation is \cite{freed}. In this
paper a non-abelian generalisation of antisymmetric rank 2 tensor
gauge potential---denoted $B$--- theory was formulated. The key point of that
discussion was the appearence of a \emph{flat} connection 1-form which
was defined via $B$. The first task was to try to understand
\emph{why} one actually needs a flat connection 1-form to play such a
dominant role in that theory. The reason according to our
understanding is as follows: it is a
well-known fact that connections in geometry are used to define the
notion of a \emph{parallel} propagation or \emph{covariant differentiation}. (The mathematical setting is
that of a principal bundle $P$ over a base space $M$ with structure group $G$).
The parallel propagation refers to sections of some associated vector
bundle $E$ say, to $P$. But there is one step that comes \emph{before}
that; namely, in order to define this parallel propagation  one
has to define the notion of a \emph{horizontal lift} of a \textsl{curve} on the
base space to a curve on the total space (here the word "horizontal" means that the vector field tangent to
the lifted curve is horizontal). This is exactly the reason why the connection
is a \emph{1-form}, namely it is the Poincare dual of a curve which is
a 1-chain.

It is then clear that if we are to introduce \emph{2-forms} as gauge
potentials and if the recepie that gauge potentials correspond to a
way of defining parallel propagation holds, then one can see that in
this case 2-forms would correspond to parallel propagation of sections
of an associated vector bundle \emph{along a surface}. This is indeed
the case in string theory for instance where the surface is the
worlsheet replacing the notion of paths, curves, of point particles. In principal bundle language this means that we would like
to have \emph{horizontal lifts of surfaces} (a surface is a 2-chain).

Now in the ordinary case of 1-chains we know from
elementary 
theory of ODE's that a horizontal lift of a curve always exists (but
it need not be unique). For a
2-chain however, such a statement---i.e. existence of a horizontal lift--- does not follow automatically. The
reason is that now we are dealing with PDE's because we have to
consider a 2-parameter family of transformations on the base manifold,
namely the flows of \emph{two} vector fields which are the vector
fields which generate the surface we would like to lift
horizontally. One then needs an \emph{integrability} condition 
for this PDE to have a solution (and hence for the horizontal lift to
exist). This is due to the fact that the horizontal lifts of each of
the vector fields (which generate the surface we want to lift) definitely \textsl{exist} but \emph{they may not form a surface} on
the total space. The necessary condition then for this to happen is that the (Lie algebra)
commutator of these two horizontal vector fields must also be a
horizontal vector field. In general this is not true, unless the
connection 1-form we used to lift them separately is \emph{flat}.\\

So we see why to each gauge potential 2-form there corresponds a
fundamental \emph{flat connection 1-form} which is naturally attached
to it.\\
 
What happens then if we would like to deal with \emph{p-branes}? Then
one has to use $(p+1)$-forms as gauge potentials because we now want
to lift horizontally a $(p+1)$-chain. We know from the
theory of integrable systems that the condition of this 
connection 1-form (used to lift separately each one of the vector
fields which generate the $(p+1)$-chain we want to lift) being flat is enough to guarantee the existence of a solution
to the relevant PDE which in turn means that the horizontal lift of this
$(p+1)$-chain  exists. (One has to be a little more careful with the
holonomy but this does not affect our argument).

We learn therefore from the above discussion that if one wants to keep
for higher degree gauge potentials the role that connection 1-forms
play in the case of point particles, there must be an underlying \emph{flat} connection 1-form
which is the fundamental object. In \cite{freed} the relation between
this flat connection 1-form and the gauge potential $B$ for the
case of the base manifold being 4-dimensional Minkowski space was
given explicitly.\\

This oughts to be true for an additional reason; as pointed out in
\cite{freed}, it is exactly due to the existence of this flat connection 1-form that
the interacting $B$-theory is (classically) equivalent to a non-linear
$\sigma $ model. And we also know that all theories involving extended
objects \emph{are}, or can be thought of as, in fact $\sigma $ models (at least in their bosonic sector
if we want to assume supersymmetry). The only difference then with strings
is that the source space will be---for a theory of $p$-branes---a
Riemannian $(p+1)$-dimensional manifold instead of just a Riemann
surface. So on top of our previous geometric discussion, flat
connection 1-forms are fundamental also for physical reasons: they are
reminiscents of the fact that our theory is in fact a $\sigma $
model. At this point we would also like to recall \cite{polyakov} in
which Polyakov rewrites $\sigma $ model Lagrangians using flat
connection 1-forms (if the source
space is not simply-connected one has to be a little more careful as
whether all flat connection 1-forms can be obtained in this way but
this is not important for our immediate discussion).\\

We now pass on to M-theory. This is supposed to be a supersymmetric non-perturbative
theory on an \emph{11-dimensional} manifold (see \cite{vafa} for a
good introduction). The crucial part is the \emph{soliton}
part of this theory since dualities can take care of the rest. Let us
assume that we start looking for a suitable Lagrangian density for, at
least to begin with, the bosonic sector of this theory. Moreover our
discussion will be restricted only to the classical level.

When one seeks for a non-perturbative theory Lagrangian---recall that
we are interested in the soliton part of the theory--- the most
sensible place to start looking at is \emph{characteristic classes}
because they are quite "stable" as being topological. But the first
difficulty comes forth immediately: we need an 11-form since the space
is 11-dimensional for a Lagrangian density and all characteristic
classes we know of in usual Chern-Weil theory are of \emph{even}
degree. But this is not the case for \emph{characteristic classes for
flat foliations of bundles} bearing in mind our previous discussion for flat
connection 1-forms as being the really fundamental objects for theories
with extended objects. In fact, characteristic classes for flat
foliations of bundles exist \textsl{only} in \emph{odd dimensional
cohomology} \cite{kamber}. What is even more exciting is that unlike
the usual characteristic classes of Chern-Weil theory they are not
defined using the curvature of a connection 1-form but \emph{flat
connection 1-forms themselves}. These are  definitely  interesting
coincidences, at least for the moment. (\emph{Note:} For \textsl{arbitrary}
foliations, characteristic classes in \textsl{any} dimension may
occur; they are "combinations" of ordinary characteristic
classes---Chern-Weil theory---and of the characteristic classes for
flat foliations of bundles mentioned above; this comment refers to F-theory).

Well, we then just go to the relevant theory (also related to
Gelfand-Fuchs cohomology and to graph cohomology---the later in
3-dimensions---) of characteristic classes for foliations, in the
special case in hand of flat foliations of bundles and we just pick up the 11th-dimensional class which is just

$$\Delta _{*}(y_6)\in H^{11}_{DR}(M)$$
where $M$ is our 11-dimensional base manifold, $H^{11}_{DR}(M)$ is its
11th-dimensional de Rham cohomology group and $y_6$ is the 6th Samelson
generator  

$$y_{6}:=\sigma\tilde{c_6}$$
where $\tilde{c_6}$ is the 6th elementary symmetric function $c_6$
multiplied by a factor $(i)^6$. The other maps $\Delta _{*}$ and
$\sigma $ appearing in the
formula are explained in detail in, for example, \cite{kamber}. This is
moreorless standard material in the geometry of foliations; the map
$\Delta _{*}$ is roughly speaking the analogue of the Weil
homomorphism and the map $\sigma $ is the map induced in cohomology
level by the universal transgression operator.\\

The last point we would like to mention is actually \emph{conjectural}: in \cite{z}
 an invariant was introduced for arbitrary foliations using the \emph{even} pairing
between  \emph{K-homology} and \emph{cyclic cohomology}. This number then we
conjecture that should be very closely related to the \emph{topological charge}
for the above mentioned Lagrangian density.

\end{document}